Manuscript type:  Empirical Article

Word Count: 4489

Acknowledgements:  none

**Collaborative Intelligence:**

**Topic Modelling of Large Language Model use in Live Cybersecurity Operations**

Dr. Martin Lochner (1, 2)
Security Data Analyst / Research Associate

Dr. Keegan Keplinger (3)
Senior Threat Intelligence Researcher

eSentire (1)
University of Waterloo (2)
SpyCloud (3)

October 20, 2024

**Abstract Page**

   **Objective:** This work describes the topic modelling of Security Operations Centre (SOC) use of a large language model (LLM), during live security operations. The goal is to better understand how these specialists voluntarily use this tool.

   **Background:** Human - automation teams have been extensively studied, but transformer-based language models have sparked a new wave of collaboration. SOC personnel at a major cybersecurity provider used an LLM to support live security operations. This study examines how these specialists incorporated the LLM into their work.

   **Method:** Our data set is the result of 10 months of SOC operators accessing GPT-4 over an internally deployed HTTP-based chat application. We performed two topic modelling exercises, first using the established BERTopic model (Grootendorst, 2022), and second, using a novel topic modeling workflow.

   **Results:** Both the BERTopic analysis and novel modelling approach revealed that SOC operators primarily used the LLM to facilitate their understanding of complex text strings. Variations on this use-case accounted for ~40% of SOC LLM usage.

   **Conclusion:** SOC operators are required to rapidly interpret complex commands and similar information. Their natural tendency to leverage LLMs to support this activity, indicates that their work-flow can be supported and augmented by designing collaborative LLM tools for use in the SOC.

   **Application:** This work can aid in creating next-generation tools for Security Operations Centres. By understanding common use-cases, we can develop workflows supporting SOC task flow. One example is a right-click context menu for executing a command line analysis LLM call directly in the SOC environment.

**Keywords**: Cybersecurity, Artificial Intelligence, Trust in Automation, Human-Automation interaction, Decision Making

*Precis: This manuscript explores the integration of a large language model (LLM) into Security Operations Centres (SOCs) during live security operations. Through topic modeling, it reveals how specialists utilize the LLM to interpret complex text, suggesting potential for developing collaborative LLM tools to enhance SOC workflows and efficiency.*

**Collaborative Intelligence: Topic Modelling of Large Language Model use in Live Cybersecurity Operations**

In this paper we investigate how Security Operations Centre (SOC) personnel, consisting of 45 operators in two centres (Ireland, Canada), use Large Language Models (LLMs) to carry out security operations tasks.  We conduct two topic modelling exercises on the SOC LLM data (for a review of topic modelling, please see Abdelrazek et al., 2023).  First, we employ an established 'BERTtopic' transformer-based approach (Grootendorst, 2022), and second, a novel approach to LLM-based topic modelling using a 'multi-shot' approach whereby we first extract topics from the dataset, and subsequently map individual LLM interactions to the shortlist of derived topics.  We will outline this process in detail and discuss the most common use-cases for which SOC operators leverage AI assistance.

**Security Operations and Collaborative Intelligence**

Security Operations Center analysts process data from multiple sources simultaneously, making critical judgments about the state of an organization's network, and in the case of a managed security provider, across the state of many organizations.  A typical SOC includes a team of cybersecurity specialists whose job is to rapidly detect, prevent, address, and remediate critical security breaches for their respective clients. The job is high-pressure and requires a broad knowledge of information technology topics in a security context.  The SOC operator uses these skills to identify, investigate, and neutralize threats. In this competitive and increasingly vital industry, any technology which improves SOC functionality will be valuable.  With the recent advancements in AI language processing, and specifically the proven abilities of these systems to interpret computer languages and other information such as network packets, threat detection rules, command line executions, and artifacts such as obfuscated code, as well as to facilitate solutions for a wide variety of security problems, the goal of human-AI collaboration in high-pressure applied contexts is becoming practical.  Human – Automation collaboration has been studied under a broad umbrella of terminology.  In this paper we will refer to human-AI teams using the term "Collaborative Intelligence".

## Human-Automation Teams in the age of LLMs

While the current trend of AI adoption may lead one to assume that human-AI collaboration arrived without precursor in Sept 2022 with the public release of GPT-3.5, the basic concepts (and their related problems and potential) have been in the public domain since the dawn of the information age, as far back as the early work in cybernetics, and the pioneering work on human-automation teams that was carried out in the 1960s and 1970s (e.g., Ferrell & Sheridan, 1967; Sheridan, 1975).  While the systems have become more advanced, the basic formulation of how humans work alongside automated systems, how failures occur in this context, and how trust is established between humans and automated systems, is largely applicable to our modern implementation of language-based AI systems (i.e. Large Language Models such as OpenAI's GPT 4).

In many ways, the current era of AI integration hints at the long-sought goal of the classical AI days (e.g., Good Old-Fashioned AI) – a coherent natural language system that can appear to reason and aggregate massive amounts of data.  We know from models of trust in automation (e.g., Lee & See, 2004) that human trust in an AI will build over time, barring critical failures which would disrupt the trust-forming process.  One of the reasons that the transformer era of human-AI collaboration is so successful, is that the current systems are good enough to instill trust in the users, such that the user can reasonably assume the system will produce accurate and useful output. In this environment, usage increases. What follows in the present work is a detailed analysis of *how* expert users, who have developed sufficient trust in an AI system, employ the LLM in the complex task of cybersecurity operations.

## Topic Modelling

'Topic Modelling' is the general term for a well-established corpus of techniques and procedures to extract topics, or meaningful labels for unorganized or unstructured data sources. For example, a researcher might have a dataset including 100 conversations from a social media site, and wish to know what was being discussed. In this example, he or she could employ any of a number of established topic modelling procedures to iteratively review all conversations, and extract topics based on the frequency of the terms included, as well as more sophisticated approaches which use pre-trained models of common topic-word relations (e.g., Grootendorst, 2022).

Topic modelling offers a way to further explore and understand large data sets beyond traditional keyword-focused methods, having applications in areas like text mining, information retrieval, and decision support systems. It can help recognize patterns in sentiment analysis and in classifying documents into different themes for easier retrieval. This enables efficient search of information, quick overviews of large text data, or identifying trends over time within textual data sets.

## Experiment 1:  SOC LLM Usage - Modelling with BERTopic

For a first pass at topic modelling of SOC LLM usage, we turned to the BERTopic repository (https://maartengr.github.io/BERTopic/index.html), which is attractive because of its simple implementation and powerful and well-known topic modelling capabilities (Egger & Yu, 2022; Raman et al., 2024).  BERTopic, developed as an open-source software library, is a topic modeling tool that uses transformer-based language models to produce a list of likely topics from a corpus of text.  The primary component driving BERTopic is BERT (Bidirectional Encoder Representations from Transformers), a language model that provides rich semantic representations of text. BERTopic uses these representations to identify and extract coherent and distinct topics from a collection of documents. It represents these topics as clusters of similar texts and allows for *topic reduction*, a feature that allows users to merge closely related topics. For more details on BERTopic, please refer to (Grootendorst, 2022).  In the present study, we use a basic implementation of this technique to extract categorical information from SOC operator LLM requests, describing the way in which SOC operators use LLMs in live security operations.

## Method

Dataset Overview

The heart of this publication, which allows us to investigate this interesting research question, is the extensive record of day-to-day LLM use by SOC operators during live security investigations, made available by eSentire, a large Canadian cybersecurity provider.  To support security operations, in September 2023 the company enabled a custom LLM application leveraging OpenAI's GPT series models, which the SOC operator could access through an internal web-based UI (Figure 1). This security architecture is also available for public usage at open source repository ([https://github.com/eSentire-Labs/LLM-Gateway](https://github.com/eSentire-Labs/LLM-Gateway)), see Figure 2. This dataset at the time of the present analysis consisted of 3787 individual SOC-LLM interactions, by 45 individual operators, at two geographic locations (Canada, Ireland). The interactions occurred over a 10 month period, between October 2023 and February 2024, with an average of 14 calls per day, with a positive skew such that more interactions were seen later in the response period (see Figure 3 for a histogram of daily interaction count over the data collection period).

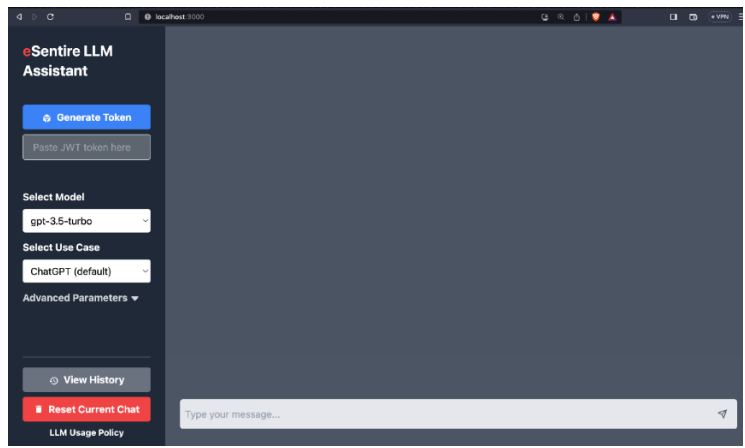

*Figure 1: Internally developed interface for LLM chat applications*

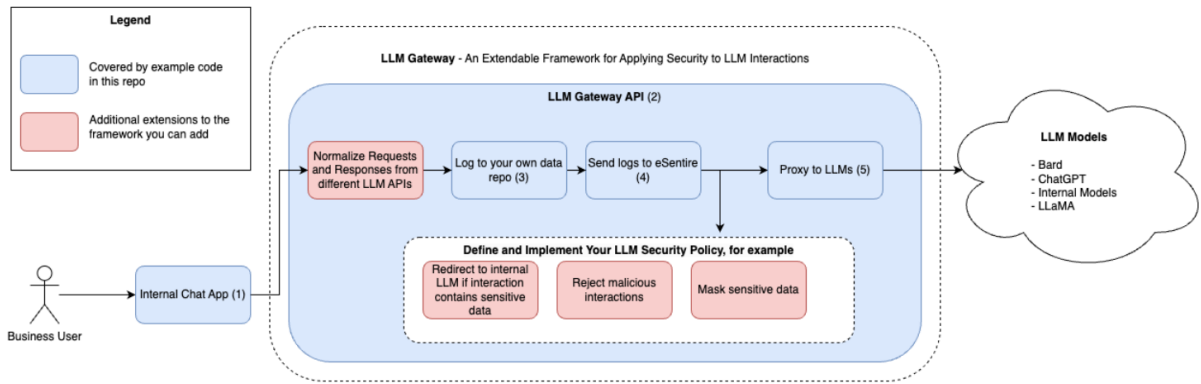

*Figure 2: eSentire's open source LLM Gateway architecture, as released in August 2023.*

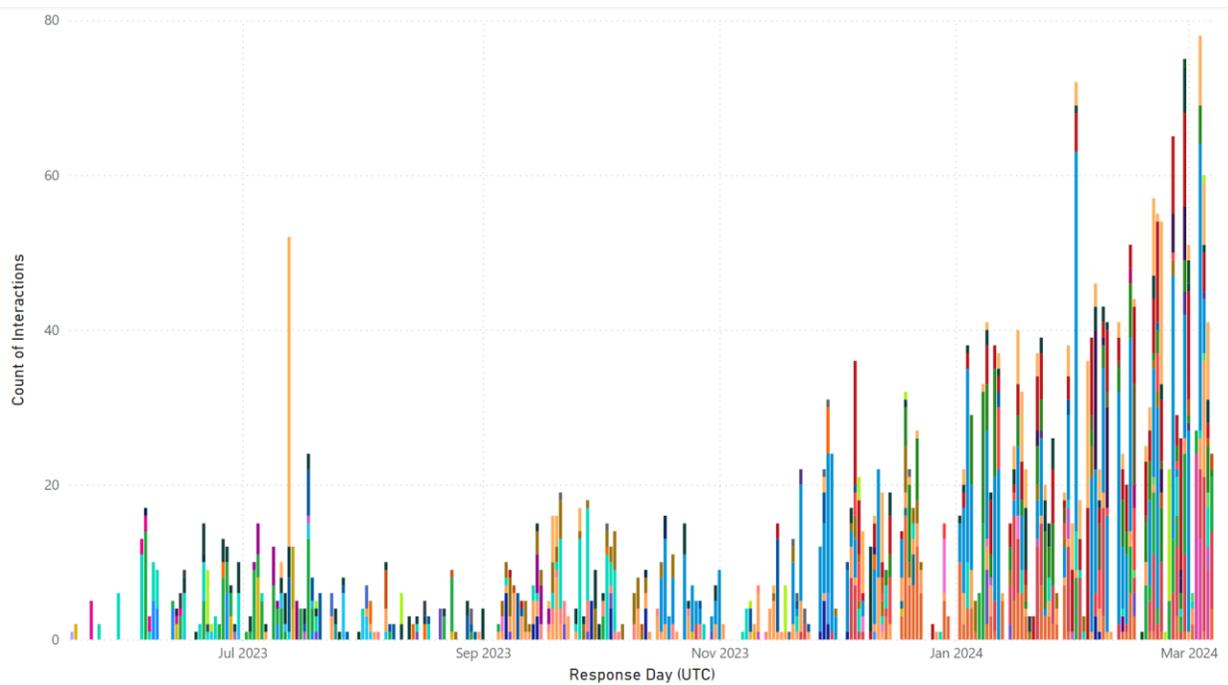

*Figure 3: Day-by-day count of interactions between SOC operators and the GPT-4 LLM application as logged by our internal LLM security gateway*

Dataset

The dataset analyzed in this study consisted of usage logs from 45 security operations professionals. The specific ages and genders of these operators is not available for publication, but they can be assumed to be in the 20 to 50 year age range. (These are generally mid-career employees in Canada and Ireland). They generally worked 8-hour shifts and were responsible as a group for 24/7 security operations, working in teams of 20 - 30 people. No live data collection was carried out, rather we analyzed previously released usage logs. All users of the internally deployed LLM signed company policy statements that indicated such data were subject to investigation and reporting. All users are anonymized, and no actual user prompts are included in this paper, excepting a single anonymous quote on page 22. This research complied with the American Psychological Association Code of Ethics.

Materials

SOC operators used the standard tools that are normally provided in their operating environment, such as incident dashboards, IP lookup services, endpoint investigation portals, alerting systems, threat intelligence platforms, filtering systems, and other proprietary software. In addition, they were provided browser-based access to an internally hosted OpenAI (GPT 4) platform, which was modelled after the standard GPT-Chat interface (Figure 1), with monitoring for compliance and data security considerations. Operators were free to navigate to the LLM system and use it in an unstructured manner to support their goals.

Design

This study involved non-structured use of the LLM during regular security operations. The SOC operators were given unrestricted access to a web-based instance of Chat GPT. While they were able to select version 3.5 or version 4, or an internally deployed model as desired, we focus here on GPT-4 due to its considerable accuracy with information such as network telemetry and computer languages.  Casework arrived over the internal SOC dashboard, and reporting and remediative actions were carried out using both proprietary and commercially available security tooling.

Data Analysis

The data were analyzed in a Jupyter notebook, using packages *asyncio, aiohttp, csv, json,* and *os*.  All higher-order mathematical operations were implemented within the BERTopic framework.  We transformed the SOC LLM calls into high-dimensional vectors ("embeddings"), using the *'all-mpnet-base-v2'* encoding model from the SentenceTransformers python library (Reimers & Gurevych, 2019).  For a recent overview of sentence encoding models, please see Kashyap et al., 2024. These embeddings effectively capture the semantic meaning of each document, which allows for their relative similarities and differences to be analyzed in a comprehensive and meaningful manner.

Results

The full set of 3787 SOC LLM calls was analyzed, resulting in a set of 90 summary categorizations and respective weights.  To get an idea of the form that these topic categorizations take, please see Table 1 below.

*Table 1: Top 12 categories for SOC LLM usage during live security investigation, as extracted by a local implementation of the BERTopic modelling library*

```
Topic 0: powershell get system object
Topic 1: command does do what
Topic 2: has file esentire detected
Topic 3: user authentication we access
Topic 4: name value rule false
Topic 5: x00 decode ncxv 101
Topic 6: summarize sentence please summary
Topic 7: cmd exe command system32
Topic 8: malicious why abused appear
Topic 9: rundll32 dll shell32 system32
Topic 10: reg software reg_dword registry
Topic 11: tcp dpkt http pcap
```

The extraction process resulted in word-group categorizations based on common word frequency in the corpus as a whole.  Each of the individual 3787 LLM calls is then assigned to one of these 90 categories, and given a weighting to indicate how well it aligns with the given topic categorization.  In this way, the BERTopic system can rapidly scan through large data sets, and provide summary level of understanding.  Subsequent steps, such as manually inferring what each of these word collections refer to by assessing the original LLM interactions, are required for a complete semantic understanding of the topics.  For example Topic 0: '`powershell executionpolicy bypass invokedexpression`' consists mainly of complex powershell commands, with user requests for interpretation of the command.  Topic 1: '`command does what do`', is similar, with users mainly requesting more information on command lines.

Combined, nearly 14% of all SOC LLM usage fell into these categories. Explanation of other topic categories continues in the General Discussion below.

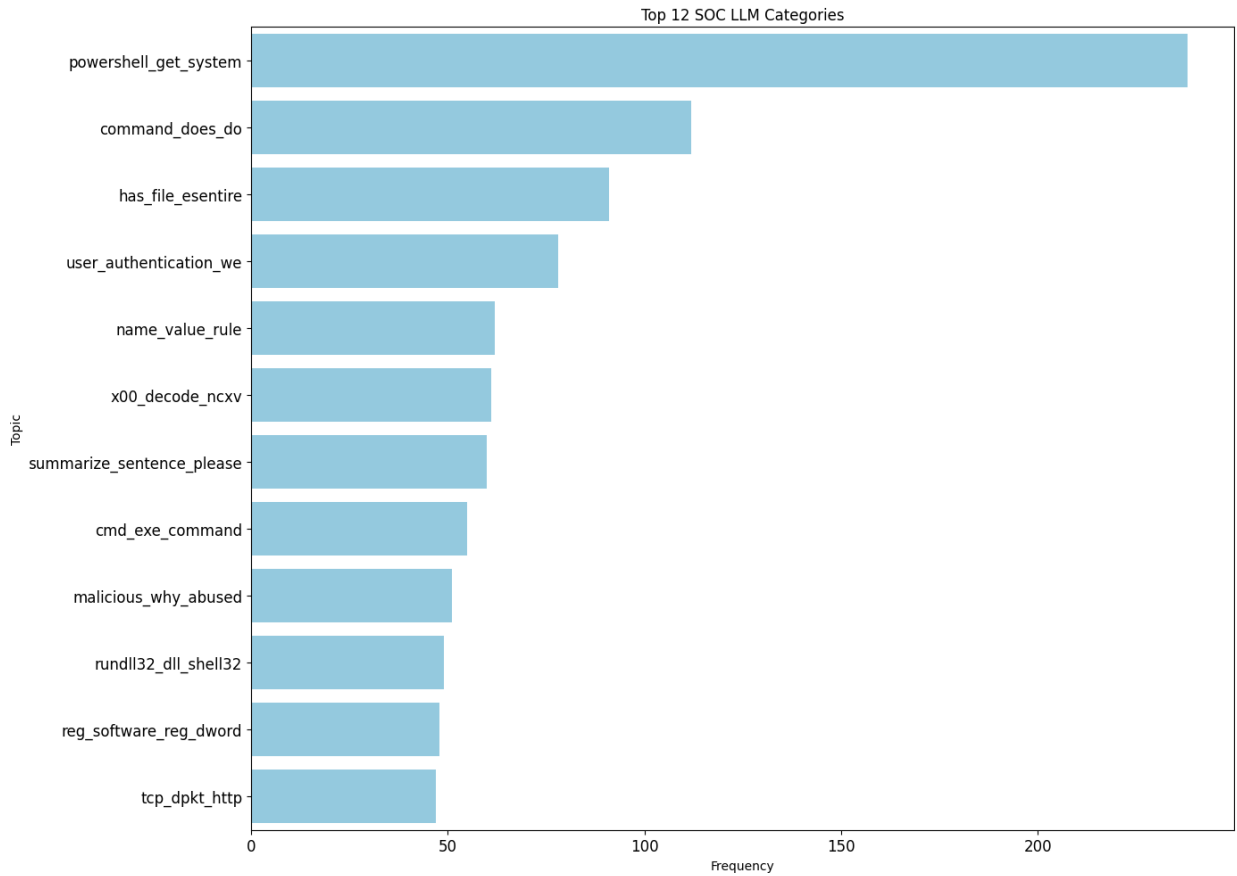

*Figure 4: Frequencies of Top 12 BERTopic classifications*

Further inspection of the weight for each term within the overall category cluster reveals the strength of their association. This analysis helps in understanding the significance of each term's contribution to the underlying topics, highlighting which terms are most representative and influential within each topic cluster. Figure 5 visualizes these weights, allowing us to see at a glance which terms carry the most weight and thus are most descriptive of the topics. We illustrate this breakdown across the same top 12 topics.

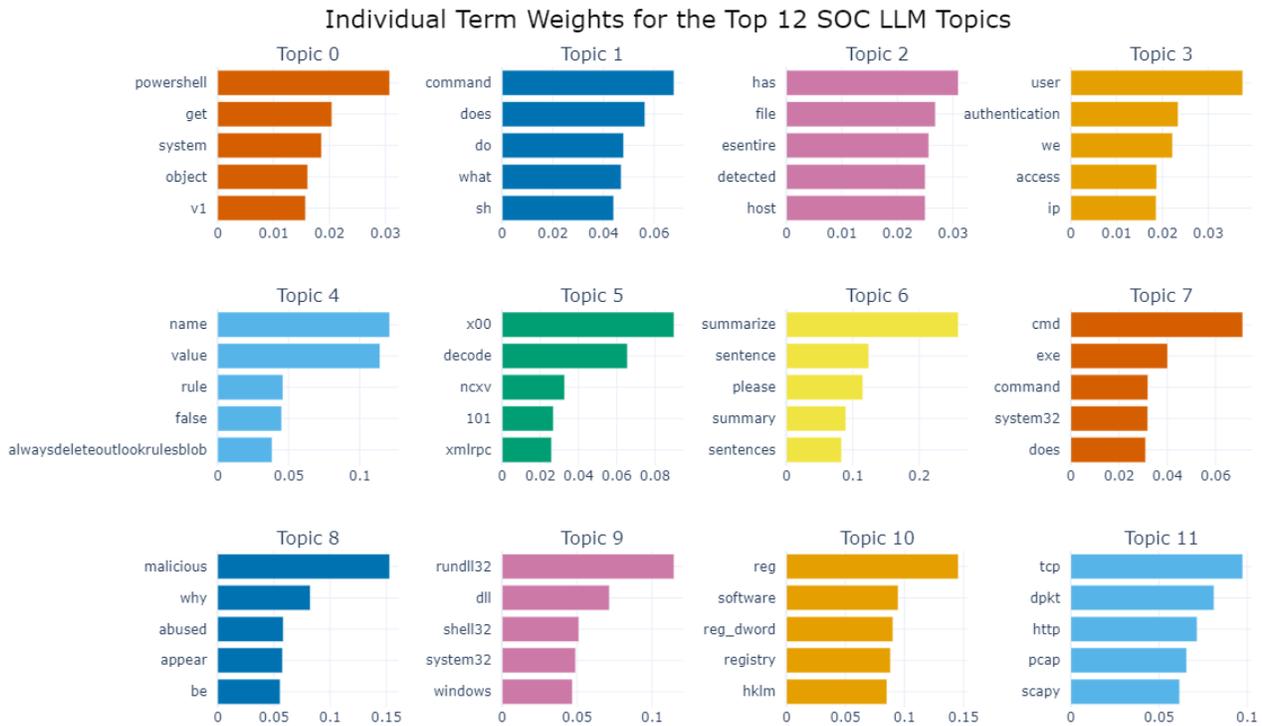

*Figure 5: Individual weights for each word within the Top 12 SOC LLM use-cases (BERTopic clusters).*

High-level clustering

The original 90 categories identified by the BERTopic model can be organized into higher-level clusters based on their similarities. In the first step, described above, individual messages are analyzed to determine common topic words using the BERTopic algorithm. These topics can then be grouped into several larger clusters, called granular clusters, by examining how closely related they are to one another. To characterize each of these higher-level clusters, representative words are identified. Similar to the initial topic word extraction, this process involves extracting the key terms most frequently associated with the messages in each cluster. By analyzing these terms, we summarize the main themes or ideas that define each higher-level cluster. Finally, these representative words are listed to provide a clear understanding of the predominant subjects within each grouped cluster. This hierarchical approach allows us to see broader themes and relationships among the initial categories, giving us a more structured and comprehensive understanding of the underlying topics.

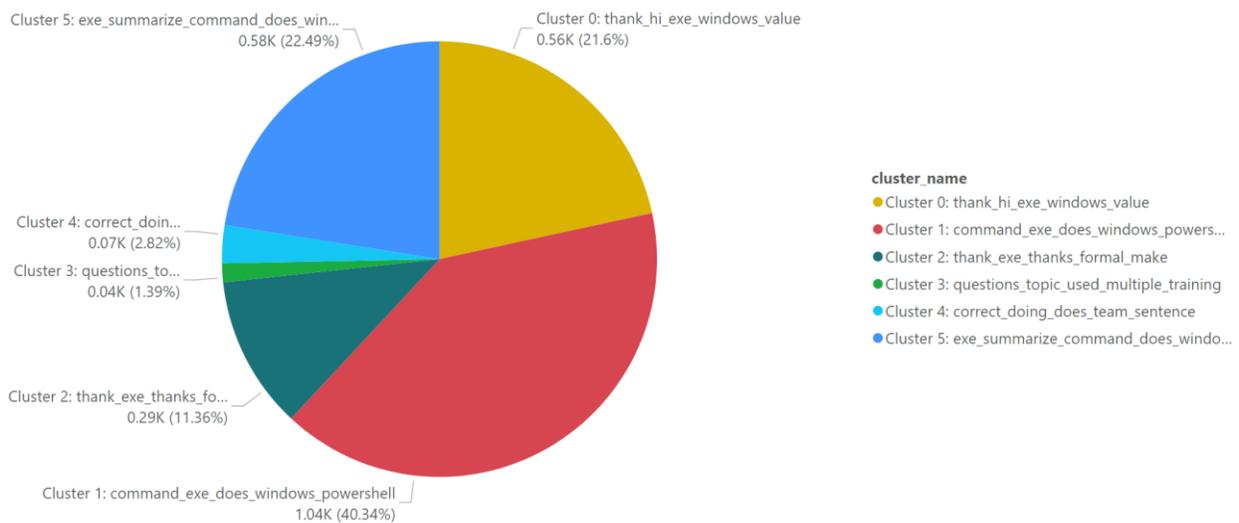

*Figure 6: Granular Clusters within the 90 BERTopic classifications, and their frequency distribution*

Discussion

This first pass at topic modelling of SOC operators LLM usage is interesting, and provides a basis for understanding how these expert workers collaborate with a language-based AI in their task goals. Towards this primary goal, we can see clearly that the "`powershell get system object`" and "`command does do what`" are by far the largest topic categories. As discussed above, these are very similar in that they both represent operators requesting help understanding a command line. Combined, this is the largest use case, representing nearly 14% of the total LLM calls.

The weak point in the present analysis is that we are left to perform a basic semantic analysis on these collections of common terms. In some cases, this can be inferred from the selected topic words, but for a better understanding, we can sort through the actual interactions with the LLM and try to infer the meaning of the common topic words. Such an inspection confirms our understanding that Topic 1 is indeed SOC analyst requests for interpretation of powershell commands. Topic 2 is a collection of other command line interpretation request, e.g., shell scripts (sh commands) and other commands. Interestingly, the third most frequent topic is "`has file esentire detected`". Closer inspection of this topic reveals that this category mainly consists of analyst requests for help in composing or rephrasing threat reports, leveraging the LLM's ability to produce elegant paragraphs from basic event details. The next topic, "`user authentication we access`" appears to be similar, in that there are many requests for sentence generation and re-phrasing, interspliced with questions about authentication methods. While it is possible to go through each of the 90 categories and assign semantic labels accordingly, we do not do so in the present analysis.

The final step above, in which we group the 90 identified topics into 6 higher-order clusters, is worth a brief discussion here as well. The resulting 6 cluster topic-words are shown in Table 2 below, in order of their frequency.

*Table 2: High-order BERTopic clusters*

```
Cluster Name                                        Count   Percentage  Inferred use case
Cluster 1: command_exe_does_windows_powershell       1044      40.34%   command line analysis
Cluster 5: exe_summarize_command_does_windows         582      22.49%   threat summaries
Cluster 0: thank_hi_exe_windows_value                 559      21.60%   mixed command analysis
Cluster 2: thank_exe_thanks_formal_make               294      11.36%   mixed analysis & rephrase
Cluster 4: correct_doing_does_team_sentence            73       2.82%   rephrase message content
Cluster 3: questions_topic_used_multiple_training      36       1.39%   content generation
```

If we dig into the individual LLM interactions within these larger clusters, it is possible to infer, in general terms, the main content of each category. In practice, there is a lot of overlap, for example, Cluster 1 is primarily made up of requests for command line analysis, but has a few instances of content generation and threat summary. Likewise, command line analysis shows up in many of the other categories. In this sense, categorization according to word frequency is a useful technique, but does have some limitations and drawbacks. Specifically, use cases are sometimes split between the larger clusters, and mapping the word-categories to an actual semantic description of the category is still a subjective and time-consuming process.

This analysis is a good starting point, and shows the power of the existing topic-modelling techniques such as BERTopic. It is very likely that a more sophisticated application of this topic modelling library would produce more fine-grained analysis and likely more accurate topic-word mapping for the individual categories. This is, however, beyond the scope of the current work; instead, we turn our focus to applying the LLMs themselves to the task of topic modelling. Given the power of the LLM tool in interpreting natural language, it seems very fitting to apply the LLM in analysis of these interactions. We will address this question in Experiment 2.

## Experiment 2

While the above topic modelling partially fulfills the goal of developing a semantic understanding of how SOC operators naturally use LLMs to solve tasks in their working environment, we find that the common-word identification approach to topic modelling, e.g., '*powershell-get-system-object*', used in Experiment 1, could use some improvement. Specifically, while the above modelling tells us what words, and word associations, are commonly occurring, it does not go so far as to tell us what the operator was doing, or attempting to do – rather, it leaves this final inference up to us.

With the question thus framed (how do we infer intent, from word frequency?), the logical step is to apply the LLM itself as a tool for inference generation. To leverage the powerful natural language processing capabilities of commercial LLMs, we developed a novel two-shot approach to classifying SOC operator LLM usage using GPT-4 as a topic modelling tool. Interestingly, this work was carried out in parallel with related work by researchers at the University of Sheffield, as currently available in pre-print on Arxiv. (Mu et al., 2024), which demonstrates that this approach may be a reasonable direction for improved topic modelling in general. The following methods and results section describes the outcome of this novel approach.

## Method

Materials & Design

The same materials and design as described above were used in this second attempt at SOC LMM topic modelling.

Analysis

We executed a novel python-based approach to topic modelling, using GPT-4 as the primary engine for topic extraction and classification. In general, our approach employed similar logic to standard topic modelling approaches, requiring two layers of processing of the raw LLM logs.

Topic Extraction Layer

Here, we divided the full corpus of 3787 LLM calls into 37 blocks of 100 calls (and one block of 87). For each of these blocks, we requested via API for GPT-4 to perform the following action:

> *"please review the following collection of 100 requests by a SOC operator to an AI assistant, and create 12 categories, or use-cases, each being 1 or 2 words."*

This resulted in a list of (38 * 12) = 456 use-cases, with potential for duplication, which represented a mid-level summary of how the SOC operators are using the LLM. We subsequently requested the LLM, in a single call, to reduce this set of 456 use cases to a single set of 20 final use cases. To this, we added a $21^{st}$ use case, 'other'. The exact text of this second prompt is:

> *"The following is a list of SOC operator use-cases. Please assess the full list, and come up with a short list of 20 high level use cases that best summarize the different types".*

At this stage of processing, the full corpus of SOC LLM calls was analyzed, and a set of 20 high-level use-cases had been generated. This process is visualized in the top half of Figure 7 below.

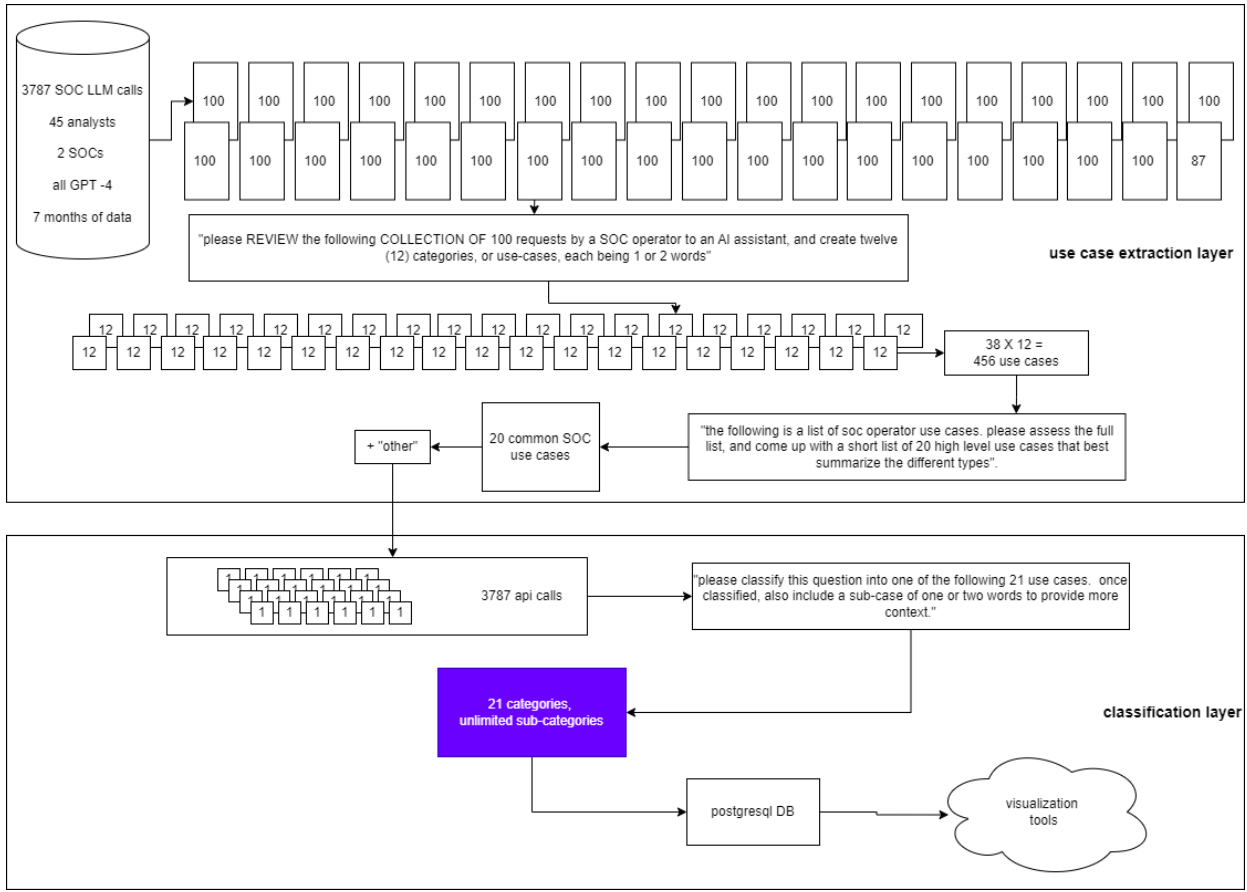

*Figure 7: Process flow diagram for python / GPT-4 based topic modelling exercise, including use-case extraction layer (top) and classification layer (bottom).*

Classification Layer

The second step in this process, represented by the lower half of Figure 7, is what we refer to as the classification layer. Here, each individual SOC operator message is submitted to the LLM, and is classified into one of the previously extracted use-cases. For example, if a SOC operator has input a command line and requested analysis of this command line, this request would be categorized into the previously extracted category of 'Command Line Analysis'. The exact text of this final instruction is as follows:

> *"Please classify this question into one of the following 21 use cases.*
>
> *Once classified, also include a sub-case of one or two words to provide*
>
> *more context".*

In this way, each of the individual SOC-LLM requests is categorized both into a primary category, which is limited to the initial set extracted in step one, and a second unconstrained sub-category, in order to provide further granularity in our analysis of how these operators are using LLMs to solve live security operations problems.

Results and Discussion

The outcome of this process is visualized in Figure 8 Figure 9.  We have demonstrated 1) a process to extract a finite set of ways in which these domain experts use the AI to support their security operations goals, and 2) a process to assign these categorical labels to each of the individual interactions.  The primary benefit of this approach, over the basic topic modelling performed in Experiment 1, is that we now can provide contextual and semantic analysis upon each interaction.  Such an approach does have known limitations, such as 'hallucinations' (e.g., Perković et al., 2024), and occasional mis-characterizations of the interaction.  However, when dealing with a large data set, such tools are very useful. Moreover, with the LLM model in use here, 'GPT-4-0613', the characterizations, or topic inferences, demonstrated are remarkably accurate.

The following graphical analysis should be useful to researchers interested generally in the technical operations of a Security Operations Centre, and specifically around how these experts naturally leverage AI assistance in performing complex tasks.  It highlights clearly the areas in which this assistance is most useful, namely, activities under the 'Command Line Operations' heading, such as command explanation, analysis, and interpretation. For example:

*'explain the following command:*
*"C:\Windows\System32\WindowsPowerShell\v1.0\powershell.exe" -noprofile -ExecutionPolicy Bypass "netsh advfirewall show allprofiles state"'*

While the operator likely has the ability to understand this command line, he or she has leveraged the LLM to produce a rapid interpretation of the command line, saving time and providing support for manual interpretation of the command features. Largely, the heavy use of LLMs within the SOC department seems to be a function of the technical complexity of their work. That is, the SOC operators find particular benefit to the LLM tool when analyzing complex commands, network packages, custom threat hunting rules, and other sophisticated computer-language analysis tasks.

*SOC LLM Usage Analysis: Primary SOC LLM usecases*

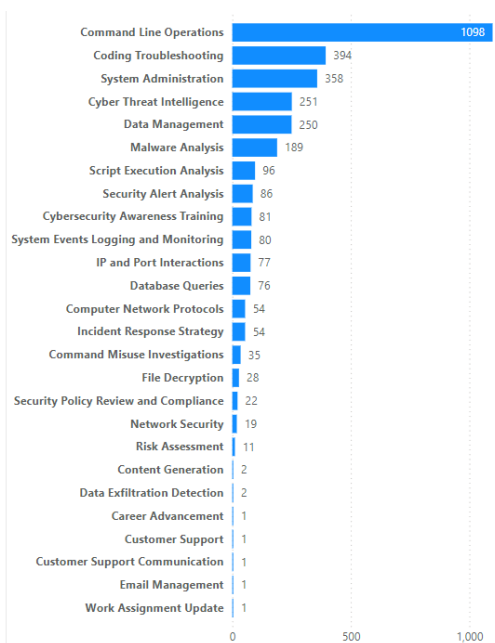

*Figure 8: Primary topic classification for SOC LLM use during live security investigations, as classified using a novel two-shot classification technique.*

Secondary categorizations for the top 8 identified primary use cases

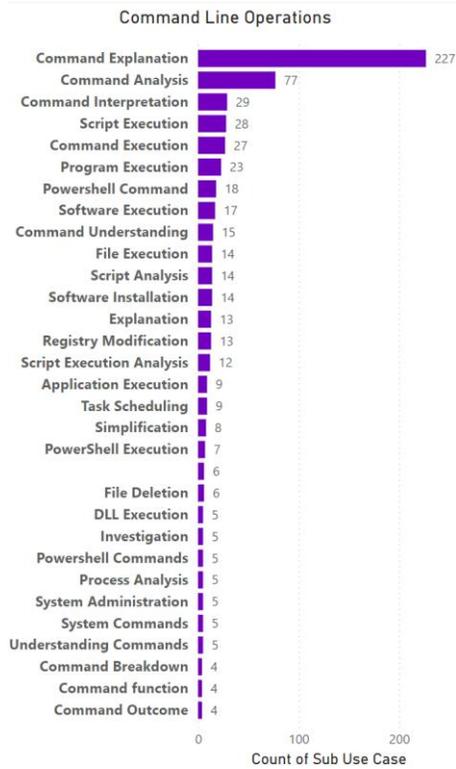
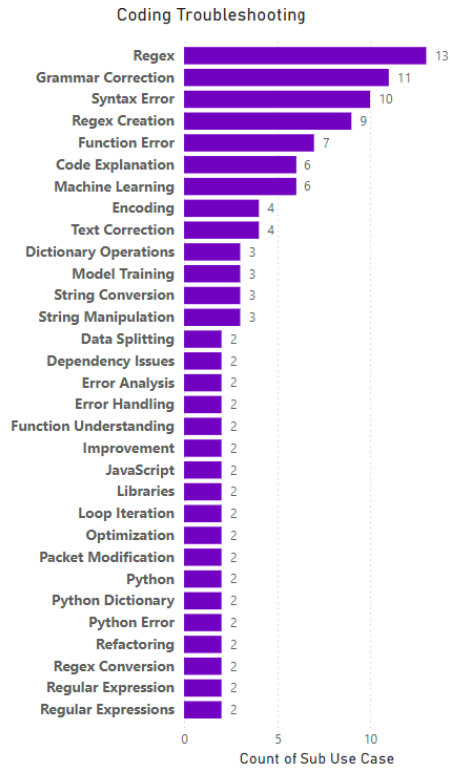
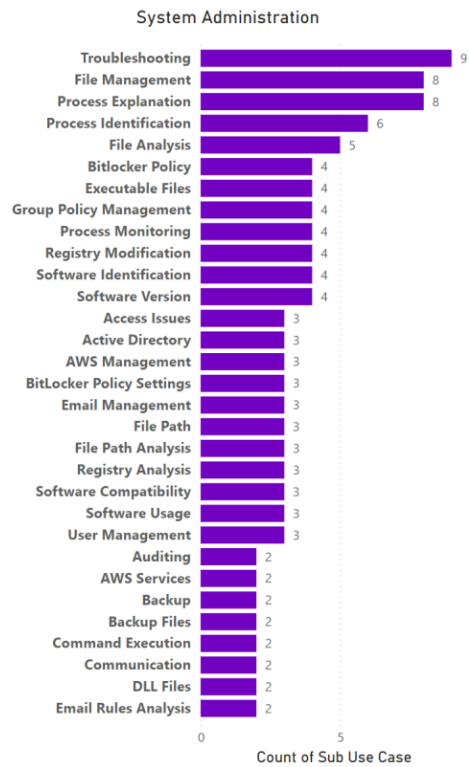
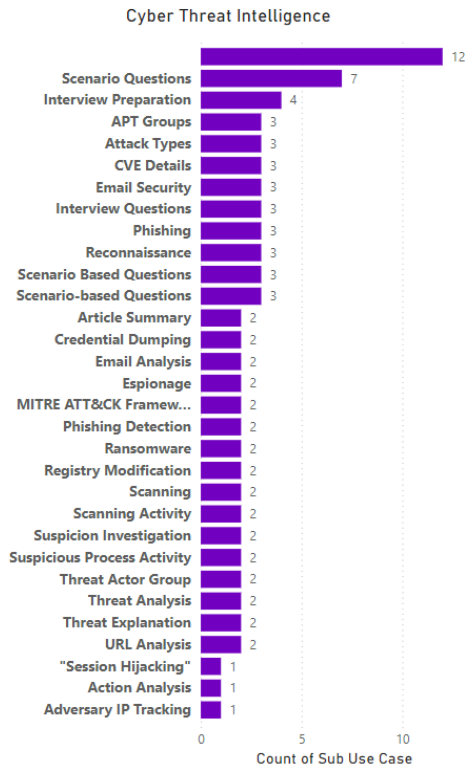

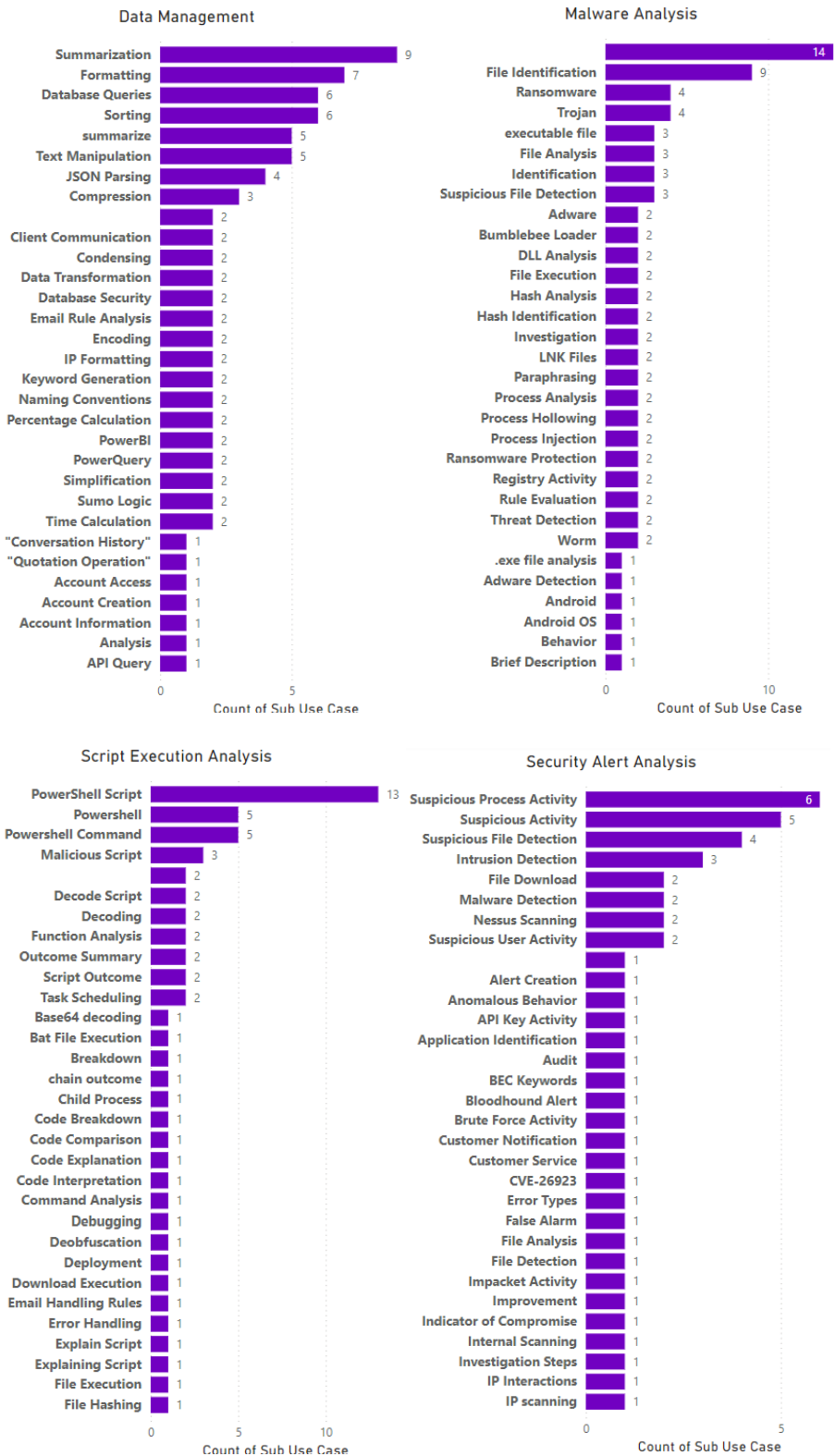

*Figure 9:Plots of sub-use-cases. Instances where the categorization directly matched the primary use case are blank, indicating that no sub-use-case was identified.*

'Other' Classification

Also of interest is the category 'Other', the 21st option in our classification layer. While the initial set of twenty use-cases was extracted from an iterative analysis of SOC LLM usage, we reasoned that there would be edge cases in which the individual call did not fit easily into one of these categories. Proportionally, the use case 'other' accounted for 517 of the classifications, or 13.6% of total responses, second only to Command Line Operations at 29%. As can be seen from inspection of Figure 10, these included messages where the SOC analyst was saying hello to the LLM, asking general questions, or providing feedback to the LLM. Individual inspection of these interactions, including the top category (the subcategory of which is blank) indicate mostly short questions, partial blocks of text or code, and casual interactions with the LLM.

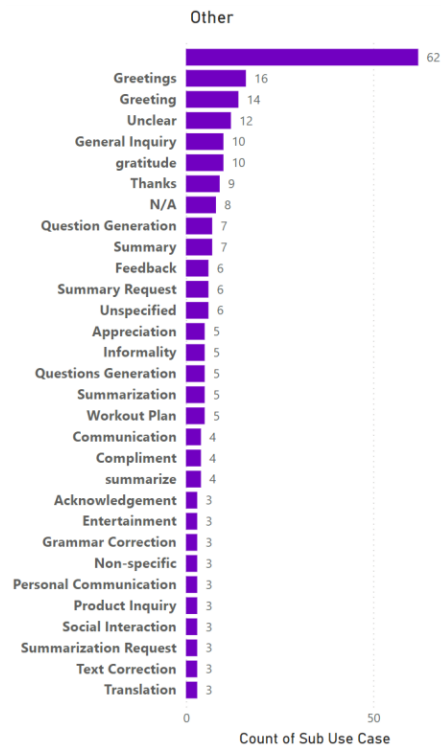

*Figure 10: 'Other' category: 21st category included to catch queries which were unrelated to the most common SOC use-cases*

General Discussion

In this paper we have 1) modelled SOC operator LLM usage with an existing topic modelling technique, BERTopic, as well as 2) performed a novel topic modelling exercise using GPT-4 as the topic modelling engine. In the first case, we generated a set of distributions across a number of topic categories, resulting in a basic understanding of how SOC operators are using these tools. However, this initial analysis leaves something to be desired – specifically, it is up to the reader to infer a reasonable semantic category for the BERTopic outcomes, e.g., "*command-fi-does-what-do*" could be interpreted as e.g., Command Line Analysis, which is indeed one of the common use-cases for SOC operators leveraging LLMs. In Experiment 2 we attempted to improve upon this initial modelling by using the more computationally sophisticated GPT-4 language model, in order to better understand how SOC operators are using these systems to support their security goals.

This novel process is interesting, and may itself provide a study into how modern LLMs can improve general topic modelling performance. Be this as it may, we believe that the main contribution of the present work to be the outcome of this topic modelling, which gives us a unique view into how security professionals collaborate with language models to support security investigations. This is interesting from a theoretical standpoint, as it provides insight into the workflow and needs of SOC operators, and also from a practical standpoint, e.g., by aiding the developers and technicians who support SOC operators, to prioritize features in the SOC tooling which would enable more efficient operations. An obvious example of this would be to take the top identified use-cases (in our example, Command Line Analysis), and begin to develop tooling enabling the operator to perform this analysis in-line, rather than cutting and pasting to a web-browser UI window. Such incremental improvements in user work-flow could lead to improvements in the overall efficiency of SOC operations, benefitting the overall cybersecurity standing of clients.

Future Directions

The present analysis gives rise to a number of interesting follow-up questions. For example, why do some operators make use of the LLM tool, while others do not? How do the results of the LLM interaction feed into the SOC operators' other tasks? The answers to such questions could further support the development of techniques to integrate these powerful tools into the standard SOC toolkit. Finally, a formal Cognitive Task Analysis of the combined LLM-SOC operations could illuminate how these tools are changing SOC workflows.

## Conclusion

Taken as a whole, the current topic modelling exercise has been very useful for understanding how these operators use the LLM in their security-oriented tasks. The results of this work have already been useful in developing novel tooling for the SOC operating environment, and could be further leveraged to develop support systems for the most common use-cases. Further, this points us in some interesting directions for future research, and gives rise to a number of practical follow-up questions.

**Key Points**

- Security Operations Centre personnel use LLMs primarily to facilitate complex command line understanding.

- This is confirmed by both an established topic modelling approach (BERTopic) and a novel LLM based approach.

- This work provides clear guidance for SOC application developers by identifying primary ways in which SOC personnel use LLMs.

RUNNING HEAD:  CYBERSECURITY LLM TOPIC MODELING                                                                28Ferrell, W. R., & Sheridan, T. B. (1967). Supervisory control of remote manipulation. *IEEE Spectrum*, *4*(10), 81–88. IEEE Spectrum. https://doi.org/10.1109/MSPEC.1967.5217126

Grootendorst, M. (2022). *BERTopic: Neural topic modeling with a class-based TF-IDF procedure* (No. arXiv:2203.05794). arXiv. http://arxiv.org/abs/2203.05794

Kashyap, A. R., Nguyen, T.-T., Schlegel, V., Winkler, S., Ng, S.-K., & Poria, S. (2024). *A Comprehensive Survey of Sentence Representations: From the BERT Epoch to the CHATGPT Era and Beyond*.

Lee, J. D., & See, K. A. (2004). Trust in automation: Designing for appropriate reliance. *Human Factors: The Journal of the Human Factors and Ergonomics Society*, *46*(1), Article 1.

McInnes, L., Healy, J., & Astels, S. (2017). hdbscan: Hierarchical density based clustering. *The Journal of Open Source Software*, *2*(11), 205. https://doi.org/10.21105/joss.00205

McInnes, L., Healy, J., & Melville, J. (2020). *UMAP: Uniform Manifold Approximation and Projection for Dimension Reduction* (No. arXiv:1802.03426). arXiv. http://arxiv.org/abs/1802.03426

Mu, Y., Dong, C., Bontcheva, K., & Song, X. (2024). *Large Language Models Offer an Alternative to the Traditional Approach of Topic Modelling* (No. arXiv:2403.16248). arXiv. http://arxiv.org/abs/2403.16248

Perković, G., Drobnjak, A., & Botički, I. (2024). Hallucinations in LLMs: Understanding and Addressing Challenges. *2024 47th MIPRO ICT and Electronics Convention (MIPRO)*, 2084–2088. https://doi.org/10.1109/MIPRO60963.2024.10569238

Raman, R., Pattnaik, D., Hughes, L., & Nedungadi, P. (2024). Unveiling the dynamics of AI applications: A review of reviews using scientometrics and BERTopic modeling. *Journal of Innovation & Knowledge*, *9*(3), 100517. https://doi.org/10.1016/j.jik.2024.100517

Appendix 1: Full list of extracted BERTopic SOC LLM categories

**[redacted due to virus scanner on upload system]**

| Topic | Top Words | Frequency |
| --- | --- | --- |
| 0 | | 238 |
| 1 | | 112 |
| 2 | | 91 |
| 3 | | 78 |
| 4 | | 62 |
| 5 | | 61 |
| 6 | | 60 |
| 7 | | 55 |
| 8 | | 51 |
| 9 | | 49 |
| 10 | | 48 |
| 11 | | 47 |
| 12 | | 45 |
| 13 | | 42 |
| 14 | | 40 |
| 15 | | 40 |
| 16 | | 39 |
| 17 | | 35 |
| 18 | | 35 |
| 19 | | 34 |
| 20 | | 34 |
| 21 | | 34 |
| 22 | | 32 |
| 23 | | 32 |
| 24 | | 29 |
| 25 | | 29 |
| 26 | | 29 |
| 27 | | 28 |
| 28 | | 28 |
| 29 | | 26 |
| 30 | | 26 |
| 31 | | 25 |
| 32 | | 25 |
| 33 | | 24 |
| 34 | | 24 |
| 35 | | 23 |
| 36 | | 23 |

| | |
|---|---|
| 37 | 23 |
| 38 | 22 |
| 39 | 22 |
| 40 | 22 |
| 41 | 22 |
| 42 | 22 |
| 43 | 22 |
| 44 | 22 |
| 45 | 21 |
| 46 | 21 |
| 47 | 21 |
| 48 | 20 |
| 49 | 19 |
| 50 | 19 |
| 51 | 19 |
| 52 | 18 |
| 53 | 18 |
| 54 | 18 |
| 55 | 18 |
| 56 | 17 |
| 57 | 17 |
| 58 | 17 |
| 59 | 17 |
| 60 | 17 |
| 61 | 16 |
| 62 | 16 |
| 63 | 16 |
| 64 | 16 |
| 65 | 16 |
| 66 | 15 |
| 67 | 15 |
| 68 | 15 |
| 69 | 15 |
| 70 | 14 |
| 71 | 14 |
| 72 | 14 |
| 73 | 14 |
| 74 | 13 |
| 75 | 13 |
| 76 | 12 |
| 77 | 12 |
| 78 | 12 |
| 79 | 12 |
| 80 | 12 |

## Author Biographies

**Dr. Martin Lochner**

Current affiliation:  (1) eSentire;   (2) University of Waterloo

Highest Degree Obtained:  Ph.D., Applied Cognitive Neuroscience, 2012, University of Guelph

Bio:
       Martin studied classical 'black box' Cognitive Psychology at University of Waterloo, Department of Psychology, specializing in cross-modal integration of sensory information in the facilitation of visual search tasks. After completing BA / MA, he completed his Ph.D. applying Multiple Object Tracking research to the driving environment at U. of Guelph.  He spent over a decade in the transportation sector, for clients such as Transport Canada, Ergonomics and Crash Avoidance Division, and Mercedes-Benz HMI Lab.  In 2017 he completed a post-doc in Cognitive Informatics with the Autonomous Systems Division at CSIRO Australia, where he led a project investigating Automation, Trust, and Human Workload.  In 2022 he started work in the Cybersecurity domain, where emerging issues surrounding the recent advancements in AI-Human interaction have evoked many of the same themes and problems that exist in the study of human trust in autonomous systems, and human – automation teams.  He currently works at eSentire, a large Managed Detection and Response provider, with the eSentire Labs team, and in collaboration with a number of interdisciplinary researchers.

**Dr. Keegan Keplinger**

Current affiliation: (1) SpyCloud

Highest Degree Obtained:  Ph.D., Mathematics and Theoretical Neuroscience, 2017, University of Waterloo

Bio:
       Keegan has degrees in physics, neuroscience, and mathematics and holds a President's Volunteer Service Award for cybersecurity. Raised a commercial fisherman out of Kodiak, Alaska, Keegan joined the cybersecurity community in 2017, as part of eSentire's Threat Intelligence team where he participated in detection engineering, tool building and the collection, analysis, and dissemination of threat intelligence. Keegan studies the linguistics and humanities of Slavic and Germanic cultures as part of a broader interest in Proto-Indo-European studies. Currently Keegan is a Senior Threat Researcher at Spycloud, and focuses on tool-development, methodology, and theory.